\documentclass[12pt,preprint]{aastex}
\usepackage{amsmath}
\usepackage{natbib}

\begin{document}
\title{Weak Gravitational Lensing of CMB - Revisit}
%\author{}

\author{B.Yu\altaffilmark{1} and T.Lu\altaffilmark{2,3}}
\altaffiltext{1} {Department of Physics, Nanjing University, Nanjing 210093, China\\ Email: yubo@chenwang.nju.edu.cn}
\altaffiltext{2} {Purple Mountain Observatory, Chinese Academy of Sciences,
Nanjing 210008, China}
\altaffiltext{3} {Joint Center for Particle, Nuclear Physics and Cosmology, Nanjing University-Purple Mountain Observatory, Nanjing 210093, China\\ Email: t.lu@mail.pmo.ac.cn}

\begin{abstract}
In this paper, we revisit the weak gravitational lensing of CMB. The
key point is that the deflection angle of CMB photon varies with the
redshift. From this we obtain a set of modified power spectra.
Compared with the ones calculated before, the temperature, the
polarization E mode and the cross power spectra correction are of
the order $0.01 \%$, while the polarization B mode correction is
about $0.4 \%$ (to $l=2000$).
\end{abstract}


\section{Introduction}
The Cosmic Microwave Background (CMB) observation is an important
probe of the cosmological physics, from which we have learnt much
about the early universe, and obtained a set of precise cosmological
parameters. However, in order to further constrain the cosmological
physics, we need more precise CMB measurements. In this case the
secondary effects begin to play an important role. The additional
CMB anisotropies due to weak gravitational lensing is such an effect
\citep{Seljak:1995ve,Zaldarriaga:1998ar,Hu:2000ee,Challinor:2005jy,Lewis:2006fu}
(and references therein).

In addition to the well-known Sachs-Wolfe effect and the integrated
Sachs-Wolfe effect, the CMB photons will be distorted by the
gravitational potential, after they decoupled at the last scattering
surface. The bend of light due to weak gravitational lensing is
small and the rms deflection is about several arcmins. But it is
important at small scales, where there is only little unlensed power
spectrum. It is also important for polarization B mode
\citep{Seljak:1996gy,Kamionkowski:1996zd}, since all the B mode
power spectrum comes from the weak gravitational lensing when the
tensor scalar ratio is zero. This is because without weak
gravitational lensing the scalar primordial perturbation does not
produce any B mode power spectrum
\citep{Zaldarriaga:1996xe,Kamionkowski:1996ks}. However, the weak
gravitational lensing will transform the polarization E mode into B
mode. The primordial tensor perturbation also produces B mode power
spectrum, whose property is important in constraining the
inflationary scenario. Observation of such primordial tensor
perturbation is regarded as a smoking gun of inflation
\citep{Mukhanov:1990me,Liddle:2000cg}, so it is very important to
subtract the B mode power spectrum induced by the weak gravitational
lensing.

As far as we know, to account for deflection angle introduced by the
weak gravitational lensing effect in the CMB power spectrum
calculation, one use the following approximation: all the photons
decouple instantaneously. In fact, not all photons are last
scattered at the same time. The visibility function has two peaks
which are respectively around the recombination era and reionization
era. About $(1-e^{-\tau})\sim 10 \%$ CMB photons are last scattered
after $z \sim 10$. For those photons last scattered at different
times, their deflection angles are different, which will affect the
weak gravitational lensing effect of CMB. Thus it affect the final
CMB power spectra, especially the polarization B mode power
spectrum, because for small tensor scalar ratio the B mode power
spectrum mainly comes from the E mode contamination induced by the
weak gravitational lensing, whereas for other types of power spectra
the weak gravitational effect is only a secondary effect.

This paper is organized as follows. In \S~\ref{sec:wgl} we provide a
theoretic framework to take into account these factors. In
\S~\ref{sec:nr} we present the numerical results. Finally in
\S~\ref{sec:dis} the discussion is given.

\section{Weak gravitational lensing of CMB}
\label{sec:wgl}

\begin{figure}[htbp]
\begin{center}
\includegraphics[width=0.6\textwidth]{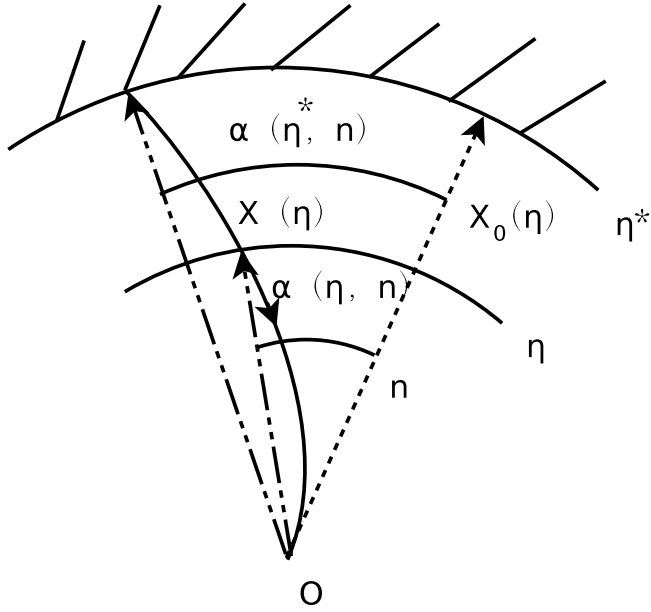}
\end{center}
\caption{Line of sight for CMB photons. The solid line is the real
line of sight $\mathbf X(\eta)$. In the unlensed case it is
approximately along the direction $-\mathbf{n}$ ($\mathbf
X_0(\eta)$). $\mathbf{\alpha} (\eta^*, \mathbf n)$ is the deflection
angle at the recombination era, while $\mathbf{\alpha}(\eta,\mathbf
n)$ is the deflection angle at a given time $\eta$.} \label{f1}
\end{figure}

For a photon last scattered at the last scattering surface, its path
of propagation is a curve. For example, it starts from the direction
$\mathbf{\hat{n}} + \mathbf{\alpha} (\eta^*, \mathbf n)$ ($\eta^*$
is the conformal time of the recombination era), propagates along
$\mathbf X$ and finally is observed in the direction $\mathbf
{\hat{n}}$ (see Fig \ref{f1}). Generally speaking, at different
times the deflection angle is different. Besides, those photons
which are last scattered at late time but are finally observed in
the direction $\mathbf {\hat{n}}$ also propagate along the same
curve after they are last scattered. The ordinary method to consider
the weak gravitational lensing effect of CMB is to set all
deflection angles $\mathbf{\alpha}(\eta,\mathbf n) \equiv
\mathbf{\alpha}(\eta^*,\mathbf n)$. Our basic idea is to do our
calculation by following the real line of sight $\mathbf X$. In this
paper only scalar primordial perturbation is discussed and other
secondary effects \citep{Zahn:2005fn,Amblard:2004ih,Santos:2003jb}
such as nonlinear evolution of the gravitational potential are not
considered.

\subsection{Temperature}

The unlensed temperature anisotropies can be represented as a line of sight integral \citep{Seljak:1996is}
\begin{equation}
T(\mathbf{\hat{n}}) = \int\,\dfrac{d^3 \mathbf{k} }{(2 \pi)^3}\,\xi({\mathbf k})\,\Delta_T(\eta_0, k, \mu),
\end{equation}
where $\xi({\mathbf k})$ is a random variable used to describe the
amplitude of the primordial fluctuation, which has the statistical
property $\langle \xi (\mathbf{k}) \xi^*(\mathbf{k}^{'}) \rangle =
(2 \pi)^3 P_{\xi}(k) \delta^3(\mathbf{k} - \mathbf{k}^{'})$,
$\Delta_T$ is
\begin{eqnarray}
\Delta_T(\eta_0, k, \mu) & = & \int_0^{\eta_0} d\eta\,{\rm e}^{-ix\mu}
    S_T(\eta,k),
\end{eqnarray}
$x=k(\eta_0-\eta)$, $\mu = \mathbf{\hat{k}} \cdot (-\mathbf{\hat{n}})$ and $S_T(\eta,k)$ is the temperature transfer function. So one obtains
\begin{eqnarray}
T(\mathbf{\hat{n}}) & = & \int_0^{\eta_0} d\eta\,\int\,\dfrac{d^3 \mathbf{k} }{(2 \pi)^3}\,\xi({\mathbf k}){\rm e}^{-ix\mu}S_T(\eta,k)     \nonumber\\
& \equiv & \int_0^{\eta_0} d\eta\,T(\eta,\mathbf X_{0}(\eta)),
\end{eqnarray}
where $\mathbf X_{0}(\eta)$ is the line of light in the unlensed case. After weak gravitational lensing effect is considered, the line of sight is not $\mathbf  X_{0}(\eta)$ but instead a new curve $\mathbf  X(\eta)$. Thus the final temperature anisotropies at direction $\mathbf{\hat{n}}$ should be
\begin{eqnarray}
\widetilde{T}(\mathbf{\hat{n}}) & = & \int_0^{\eta_0} d\eta\,T(\eta,\mathbf  X(\eta)),
\label{Tem}
\end{eqnarray}
with $T(\eta,\mathbf  X(\eta))$ being calculated as the one in the
unlensed case. It should be noted that generally speaking
$T(\eta,\mathbf  X(\eta))$ is not the one in the unlensed case,
however considering that the temperature source term is only low
pole angular dependence, one can approximately replace the real
$T(\eta,\mathbf  X(\eta))$ with the one in the unlensed case (the
same approximation is made in the polarization case). The new line
of sight $\mathbf X(\eta) = \chi (\mathbf{\hat{n}} +\mathbf{\alpha}
(\mathbf \eta, \mathbf n))$ with $\chi = \eta_{0} - \eta -2
\,\int^{\eta}_{\eta_{0}}\,d\eta^{'} \Psi$, where $\Psi$ is the 'Weyl
potential' defined by $\Psi = \frac{(\Psi_N - \Phi_N)}{2}$
($\Psi_N$, $\Phi_N$ is the gravitational potential appearing in the
line element $d^2s = a^2(\eta) \{-(1+2 \Psi_N)d^2 \eta + (1+2
\Phi_N)[d^2 \chi + f^2_K(\chi)(d^2 \theta + \sin^2 \theta d^2
\varphi)] \}$). In the following the radial delays are ignored, so
$\chi = \eta_{0} - \eta$. $\mathbf{\alpha} (\mathbf  \eta, \mathbf
n)$ is the deflection angle at time $\eta$
\begin{equation}
\mathbf{\alpha} (\eta,\mathbf{n}) = \nabla_{\mathbf{ \hat{n}}} \psi(\eta,\mathbf n)
\end{equation}
with
\begin{equation}
\psi(\eta,\mathbf n) = -2 \int_0^{\chi} d\chi^{'} \frac{f_K(\chi-\chi^{'})}{f_K(\chi)f_K(\chi^{'})} \Psi(\chi^{'} \mathbf{\hat{n}}; \eta_0 -\chi^{'}),
\label{psi}
\end{equation}
where $\nabla_{\mathbf{ \hat{n}}}$ represents the angular derivative and $f_K(\chi) = \chi$ for a flat universe \citep{Kaiser:1996tp,Lewis:2006fu}. It is obvious that $\mathbf{\alpha} (\eta, \mathbf{n})$ is different from $\mathbf{\alpha} (\eta^*, \mathbf{n})$. The ordinary method corresponds to replacing $\mathbf{\alpha} (\eta, \mathbf{n})$ by $\mathbf{\alpha} (\eta^*, \mathbf{n})$.

In the flat-sky approximation, we use the 2D Fourier transformation convention
\begin{eqnarray}
T(\eta,\mathbf X(\eta)) = T(\eta,\mathbf n +\mathbf{\alpha} (\mathbf  \eta, \mathbf n)) = \int \dfrac{d^2 \mathbf{l} }{(2 \pi)^2} T(\eta,\mathbf{l}) {\rm e}^{i \mathbf{l} \cdotp (\mathbf{\hat{n}} +\mathbf{\alpha} (\mathbf  \eta, \mathbf n))}.
\end{eqnarray}
Following the standard methods, the temperature correlator between two directions can be represented as
\begin{eqnarray}
 \widetilde{\xi}(r) & \equiv &  \langle \widetilde{T}^s(\hat{n}) \widetilde{T}^s(\hat{n^{'}}) \rangle                           \nonumber\\
& = & \int^{\eta_0}_0\, d\eta \int^{\eta_0}_0\, d\eta^{'} \int \dfrac{d^2 \mathbf{l} }{(2 \pi)^2} \int \dfrac{d^2 \mathbf{l^{'}} }{(2 \pi)^2} \langle {\rm e}^{i \mathbf l \cdotp (\mathbf{\hat{n}} + \mathbf{\alpha} (\eta, \mathbf n))}                          \nonumber\\
& & {\rm e}^{- i \mathbf{l^{'}} \cdotp (\mathbf{\hat{n}}^{'} +\mathbf{\alpha} (\eta^{'}, \mathbf n^{'}))} \rangle \langle T(\eta,\mathbf{l}) T^*(\eta^{'},\mathbf{l^{'}}) \rangle   \nonumber\\
& = & \int^{\eta_0}_0\, d\eta \int^{\eta_0}_0\, d\eta^{'} \int
\dfrac{d^2 \mathbf{l} }{(2 \pi)^2}  {\rm e}^{i \mathbf{l} \cdotp
\mathbf{r} } \langle {\rm e}^{i \mathbf{l} \cdotp (\mathbf{\alpha}
(\eta, \mathbf n)-\mathbf{\alpha} (\eta^{'}, \mathbf n^{'}))}
\rangle C^{TT(\eta \eta^{'})}_l,
\end{eqnarray}
where we have ignored the $\langle T \Psi \rangle$ term since it is
very small, and  in the last line we have already used $\langle
T(\eta,\mathbf{l}) T^*(\eta^{'},\mathbf{l^{'}}) \rangle = (2 \pi)^2
\delta^2 (\mathbf l - \mathbf l^{'})      C^{TT(\eta \eta^{'})}_l$
and $\mathbf r =\mathbf{\hat{n}} - \mathbf{\hat{n}}^{'}$. Detailed
calculation of $C^{TT(\eta \eta^{'})}_l$ is presented in Appendix B.
$\langle {\rm e}^{i \mathbf{l} \cdotp (\mathbf{\alpha} (\eta,
\mathbf n)-\mathbf{\alpha} (\eta^{'}, \mathbf n^{'}))} \rangle$ can
be calculated following the same methods as those of
\citep{Lewis:2006fu,Zaldarriaga:1998ar}, the difference is that
$\mathbf{\alpha}$ is now time dependent. For the Gaussian variate
$\mathbf{l} \cdotp (\mathbf{\alpha} (\eta, \mathbf
n)-\mathbf{\alpha} (\eta^{'}, \mathbf n^{'}))$, one has
\begin{eqnarray}
\langle {\rm e}^{i \mathbf{l} \cdotp (\mathbf{\alpha} (\eta, \mathbf n)-\mathbf{\alpha} (\eta^{'}, \mathbf n^{'}))}\rangle = \exp (-\frac{1}{2}\langle[\mathbf{l} \cdotp (\mathbf{\alpha} (\eta, \mathbf n)-\mathbf{\alpha} (\eta^{'}, \mathbf n^{'}))]^2\rangle).
\end{eqnarray}
Now one proceeds to compute the correlation tensor
\begin{eqnarray}
\langle \mathbf{\alpha}_i (\eta, \mathbf n) \mathbf{\alpha}_j (\eta^{'}, \mathbf n^{'}) \rangle & = & \langle \nabla_i \psi(\eta,\mathbf n) \nabla_j \psi(\eta^{'},\mathbf n^{'}) \rangle  \nonumber\\
& = & \int \dfrac{d^2 \mathbf{l} }{(2 \pi)^2} l_i l_j {\rm e}^{i \mathbf{l} \cdotp \mathbf{r} } C^{\psi  \psi(\eta \eta^{'})}_l
\end{eqnarray}
where we have used $\langle \psi(\eta,\mathbf l) \psi^* (\eta^{'},
\mathbf l^{'}) \rangle = (2 \pi)^2 \delta^2 (\mathbf l - \mathbf
l^{'})      C^{\psi  \psi(\eta \eta^{'})}_l $. The detailed form of
$C^{\psi  \psi(\eta \eta^{'})}_l$ is given in Appendix B. One note
that for a given $\eta$ and $\eta^{'}$, the correlation tensor can
only depend on $\delta_{ij}$ and the trace-free tensor
$\mathbf{\hat{r}}_{\langle i} \mathbf{\hat{r}}_{j \rangle} =
\mathbf{\hat{r}}_i \mathbf{\hat{r}}_j-\frac{1}{2}\delta_{ij}$.
Therefore,
\begin{eqnarray}
\langle \mathbf{\alpha}_i (\eta, \mathbf n) \mathbf{\alpha}_j
(\eta^{'}, \mathbf n^{'}) \rangle = \frac{1}{2}
C^{\eta\eta^{'}}_{gl}(r) \delta_{ij} -
C^{\eta\eta^{'}}_{gl,2}(r)\mathbf{\hat{r}}_{\langle i}
\mathbf{\hat{r}}_{j \rangle}
\end{eqnarray}
with
\begin{eqnarray}
C^{\eta\eta^{'}}_{gl}(r) & = &  \int \dfrac{d l }{2 \pi} l^3 C^{\psi  \psi(\eta \eta^{'})}_l J_0(lr)                 \nonumber\\
C^{\eta\eta^{'}}_{gl,2}(r) & = &  \int \dfrac{d l }{2 \pi} l^3 C^{\psi  \psi(\eta \eta^{'})}_l J_2(lr).
\end{eqnarray}
Thus
\begin{eqnarray}
&&\langle[\mathbf{l} \cdotp (\mathbf{\alpha} (\eta, \mathbf n)-\mathbf{\alpha} (\eta^{'}, \mathbf n^{'}))]^2\rangle           \nonumber\\
= && l_i l_j \langle (\mathbf{\alpha} (\eta, \mathbf n)-\mathbf{\alpha} (\eta^{'}, \mathbf n^{'}))_i (\mathbf{\alpha} (\eta, \mathbf n)-\mathbf{\alpha} (\eta^{'}, \mathbf n^{'}))_j \rangle        \nonumber\\
= && l^2 [\dfrac{C^{\eta\eta}_{gl}(0) + C^{\eta^{'}\eta^{'}}_{gl}(0)}{2} - C^{\eta\eta^{'}}_{gl}(r) ]+l^2 \cos 2\phi C^{\eta\eta^{'}}_{gl,2}(r),
\end{eqnarray}
where we have used the following properties: $C^{\eta\eta^{'}}_{gl,2}(0) = 0 $, since $J_2(0)=0$; $C^{\eta\eta^{'}}_{gl}(r) = C^{\eta^{'}\eta}_{gl}(r)$, $C^{\eta\eta^{'}}_{gl,2}(r) = C^{\eta^{'}\eta}_{gl,2}(r)$, since $C^{\psi  \psi(\eta \eta^{'})}_l$ is symmetrical in the time variable and $\phi$ is the angle between $\mathbf{l}$ and $\mathbf{r}$. Putting them together, one obtains the correlation function as follows,
\begin{eqnarray}
\widetilde{\xi}(r) & = & \int^{\eta_0}_0\, d\eta \int^{\eta_0}_0\, d\eta^{'} \int \dfrac{d^2 \mathbf{l} }{(2 \pi)^2}  {\rm e}^{i \mathbf{l} \cdotp \mathbf{r} } C^{TT(\eta \eta^{'})}_l  \nonumber\\
& & \exp (-\frac{1}{2} l^2 [ \dfrac{C^{\eta\eta}_{gl}(0) + C^{\eta^{'}\eta^{'}}_{gl}(0)}{2} - C^{\eta\eta^{'}}_{gl}(r) + \cos 2\phi C^{\eta\eta^{'}}_{gl,2}(r)  ])       \nonumber\\
& = & \int^{\eta_0}_0\, d\eta \int^{\eta_0}_0\, d\eta^{'} \int \dfrac{dl}{2 \pi} l  C^{TT(\eta \eta^{'})}_l  \nonumber\\
& & \exp (-\frac{1}{2} l^2 [ \dfrac{C^{\eta\eta}_{gl}(0) + C^{\eta^{'}\eta^{'}}_{gl}(0)}{2} - C^{\eta\eta^{'}}_{gl}(r) ])  \nonumber\\
& & [J_0(lr) + \frac{l^2}{2} C^{\eta\eta^{'}}_{gl,2}(r) J_2(lr) +\ldots ].
\end{eqnarray}
When all deflection angles $\mathbf{\alpha}(\eta,\mathbf n)$ are
replaced by the value at the last scattering surface
$\mathbf{\alpha}(\eta^*,\mathbf n)$, $\exp (-\frac{1}{2} l^2 [
\dfrac{C^{\eta\eta}_{gl}(0) + C^{\eta^{'}\eta^{'}}_{gl}(0)}{2} -
C^{\eta\eta^{'}}_{gl}(r) + \cos 2\phi C^{\eta\eta^{'}}_{gl,2}(r) ])$
becomes a constant, which can be extracted out from the time
integral. Thus $\int^{\eta_0}_0\, d\eta \int^{\eta_0}_0\, d\eta^{'}
C^{TT(\eta \eta^{'})}_l = C^T_l$. The temperature correlator
$\widetilde{\xi}(r)$ reduces to the ordinary one.

\subsection{Polarization}

As was done in \citep{Zaldarriaga:1996xe,Lewis:2006fu}, the
polarization field $P = Q +i U$ ($Q$, $U$ are the Stokes parameters)
can be decomposed into
\begin{eqnarray}
P & = & \sum_k ( E_k + i B_k ) \,_2Q_k       \nonumber\\
  & = & \sum_k S_k ( \bar{E}_k + i \bar{B}_k ) \,_2Q_k,
\end{eqnarray}
where $\,_2Q_k$ is the spin $2$ harmonics, $E_k,B_k$ are their
harmonic coefficients, and $\bar{E}_k,\bar{B}_k$ are the harmonic
coefficients of spin zero quantity $\bar{E},\bar{B}$ (corresponding
to $\tilde{E}(\hat{\mathbf{n}}),\tilde{B}(\hat{\mathbf{n}})$ defined
in \citep{Zaldarriaga:1996xe}). These two kinds of harmonic
coefficient are related by some quantity $S_k$, which expressed in
spherical harmonic basis is $S_l = [(l-1)l(l+1)(l+2)]^{-1/2}$ .
Usually $\bar{E},\bar{B}$ is the line of sight integral over time
$\bar{E} = \int_0^{\eta_0} d\eta\,\bar{E} (\eta,\mathbf n)$,
$\bar{B} = \int_0^{\eta_0} d\eta\,\bar{B} (\eta,\mathbf n)$. For
example, scalar type perturbations produce a $\bar{E}$
\citep{Lin:2004xy}
\begin{eqnarray}
\bar{E}(\mathbf n) & = & \int\,\dfrac{d^3 \mathbf{k} }{(2 \pi)^3}\,\xi({\mathbf k})\,\Delta_{P}(\eta_0, k, \mu)  \nonumber\\
& = & \int_0^{\eta_0} d\eta\,\int\,\dfrac{d^3 \mathbf{k} }{(2 \pi)^3} \xi({\mathbf k})\, \frac{3}{4} g(\eta)\,\Pi(\eta,k)
        (1+\partial_x^2)^2 \left( x^2\,{\rm e}^{-ix\mu}\right)        \nonumber\\
& \equiv & \int_0^{\eta_0} d\eta\,\bar{E}(\eta,\mathbf  X_0 (\eta))
\label{ebar}
\end{eqnarray}
and zero $\bar{B}$, where $\Delta_{P}$ is the polarization
anisotropy, $g(\eta)$ is the visibility function and $\Pi =
\Delta_{T2} + \Delta_{P2} + \Delta_{P0}$ ($\Delta_{T2}$ is the
quadrupole of $\Delta_{T}$, $\Delta_{P0(2)}$ is the monopole
(quadrupole) of $\Delta_{P}$). Thus $P$ can be rewritten as
\begin{eqnarray}
P & = & \int_0^{\eta_0} d\eta\, \sum_k s_k \bar{E}_k(\eta) \,_2Q_k        \nonumber\\
& = & \int_0^{\eta_0} d\eta\, \sum_k E_k(\eta) \,_2Q_k    \nonumber\\
& \equiv & \int_0^{\eta_0} d\eta\,P(\eta,\mathbf n),
\end{eqnarray}
where $\bar{E}_k(\eta)$ is the harmonic coefficient of $\bar{E}(\eta,\mathbf  X_0 (\eta))$ and $E_k(\eta)= s_k \bar{E}_k(\eta)$. In the flat-sky approximation, it becomes \citep{Lewis:2006fu}
\begin{eqnarray}
P(\eta,\mathbf n)  =  -\int \dfrac{d^2 \mathbf{l} }{(2 \pi)^2} E(\eta,\mathbf l) {\rm e}^{2 i \phi_{\mathbf l} } {\rm e}^{i \mathbf{l} \cdotp \mathbf{\hat{n}} },
\end{eqnarray}
where $\phi_{\mathbf l}$ is the angle between $\mathbf{l}$  and the fixed basis $\mathbf e_x$ in the flat-sky approximation. Analogous to $P$, $P^*$ can be decomposed by spin $-2$ harmonics. Then one obtains
\begin{eqnarray}
P^*(\eta,\mathbf n)  =  -\int \dfrac{d^2 \mathbf{l} }{(2 \pi)^2} E(\eta,\mathbf l) {\rm e}^{- 2 i \phi_{\mathbf l} } {\rm e}^{i \mathbf{l} \cdotp \mathbf{\hat{n}} }.
\end{eqnarray}
When weak gravitational lensing effect is considered,
\begin{eqnarray}
\widetilde{P} = \int_0^{\eta_0} d\eta\,P(\eta,\mathbf n + \mathbf{\alpha} (\eta, \mathbf n)).
\label{Pol}
\end{eqnarray}

Now one can calculate the polarization correlation function, as was
done in the standard reference \citep{Lewis:2006fu}. $\mathbf r =
\mathbf {\hat{n}} - \mathbf {\hat{n}}^{'}$ is chosen as the basic
coordinate axis to define the Stokes parameter. The polarization
field $P_r$ defined above is connected by a rotation to the original
polarization field defined in $\mathbf e_x$ basis, $P_r={\rm
e}^{-2i\phi_{\mathbf r}}P$, where $\phi_{\mathbf r}$ is the
rotational angle. Then one obtains the correlator
\begin{eqnarray}
\widetilde{\xi}_{+}(r) & \equiv &  \langle \widetilde{P}(\mathbf{\hat{n}}) \widetilde{P}(\mathbf{\hat{n}}^{'}) \rangle                           \nonumber\\
& = & \int^{\eta_0}_0\, d\eta \int^{\eta_0}_0\, d\eta^{'} \langle P (\eta,\mathbf{\hat{n}} + \mathbf{\alpha}(\eta,\mathbf n)) P (\eta^{'},\mathbf{\hat{n}}^{'} + \mathbf{\alpha}(\eta^{'},\mathbf n^{'})) \rangle          \nonumber\\
& = & \int^{\eta_0}_0\, d\eta \int^{\eta_0}_0\, d\eta^{'} \int \dfrac{d^2 \mathbf{l} }{(2 \pi)^2} \int \dfrac{d^2 \mathbf{l^{'}} }{(2 \pi)^2} \langle {\rm e}^{i \mathbf l \cdotp (\mathbf{\hat{n}} +\mathbf{\alpha} (\eta, \mathbf n))}                          \nonumber\\
& & {\rm e}^{- i \mathbf{l^{'}} \cdotp (\mathbf{\hat{n}}^{'} +\mathbf{\alpha} (\eta^{'}, \mathbf n^{'}))} \rangle {\rm e}^{2 i \phi_{\mathbf l} } {\rm e}^{-2 i \phi_{\mathbf l^{'}} } \langle E(\eta,\mathbf l) E^{*}(\eta^{'},\mathbf l^{'}) \rangle   \nonumber\\
& = & \int^{\eta_0}_0\, d\eta \int^{\eta_0}_0\, d\eta^{'} \int \dfrac{d^2 \mathbf{l} }{(2 \pi)^2}  {\rm e}^{i \mathbf{l} \cdotp \mathbf{r} } C^{EE(\eta \eta^{'})}_l  \nonumber\\
& &\exp (-\frac{1}{2} l^2 [ \dfrac{C^{\eta\eta}_{gl}(0) + C^{\eta^{'}\eta^{'}}_{gl}(0)}{2} - C^{\eta\eta^{'}}_{gl}(r) + \cos 2\phi C^{\eta\eta^{'}}_{gl,2}(r)  ])     \nonumber\\
& = & \int^{\eta_0}_0\, d\eta \int^{\eta_0}_0\, d\eta^{'} \int \dfrac{d l }{2 \pi}  l C^{EE(\eta \eta^{'})}_l  \nonumber\\
& & \exp (-\frac{1}{2} l^2 [ \dfrac{C^{\eta\eta}_{gl}(0) + C^{\eta^{'}\eta^{'}}_{gl}(0)}{2} - C^{\eta\eta^{'}}_{gl}(r) ])  \nonumber\\
& & [J_0(lr) + \frac{l^2}{2} C^{\eta\eta^{'}}_{gl,2}(r) J_2(lr) +\ldots ],
\end{eqnarray}
and
\begin{eqnarray}
\widetilde{\xi}_{-}(r) & \equiv &  \langle {\rm e}^{-4i\phi_{\mathbf r}} \widetilde{P}(\mathbf{\hat{n}}) \widetilde{P}(\mathbf{\hat{n}}^{'}) \rangle                           \nonumber\\
& = & \int^{\eta_0}_0\, d\eta \int^{\eta_0}_0\, d\eta^{'} \langle {\rm e}^{-4i\phi_{\mathbf r}} P (\eta,\mathbf{\hat{n}} + \mathbf{\alpha}(\eta,\mathbf n)) (P^* (\eta^{'},\mathbf{\hat{n}}^{'} + \mathbf{\alpha}(\eta^{'},\mathbf n^{'})))^* \rangle          \nonumber\\
& = & \int^{\eta_0}_0\, d\eta \int^{\eta_0}_0\, d\eta^{'} \int \dfrac{d^2 \mathbf{l} }{(2 \pi)^2} \int \dfrac{d^2 \mathbf{l^{'}} }{(2 \pi)^2} \langle {\rm e}^{i \mathbf l \cdotp (\mathbf{\hat{n}} +\mathbf{\alpha} (\eta, \mathbf n))}                          \nonumber\\
& & {\rm e}^{- i \mathbf{l^{'}} \cdotp (\mathbf{\hat{n}}^{'} +\mathbf{\alpha} (\eta^{'}, \mathbf n^{'}))} \rangle {\rm e}^{-4i\phi_{\mathbf r}} {\rm e}^{2 i \phi_{\mathbf l} } {\rm e}^{ 2 i \phi_{\mathbf l^{'}} } \langle E(\eta,\mathbf l)E^{*}(\eta^{'},\mathbf l^{'}) \rangle   \nonumber\\
& = & \int^{\eta_0}_0\, d\eta \int^{\eta_0}_0\, d\eta^{'} \int \dfrac{d^2 \mathbf{l} }{(2 \pi)^2}  {\rm e}^{i \mathbf{l} \cdotp \mathbf{r} } C^{EE(\eta \eta^{'})}_l {\rm e}^{4i\phi}  \nonumber\\
& &\exp (-\frac{1}{2} l^2 [ \dfrac{C^{\eta\eta}_{gl}(0) + C^{\eta^{'}\eta^{'}}_{gl}(0)}{2} - C^{\eta\eta^{'}}_{gl}(r) + \cos 2\phi C^{\eta\eta^{'}}_{gl,2}(r)  ])     \nonumber\\
& = & \int^{\eta_0}_0\, d\eta \int^{\eta_0}_0\, d\eta^{'} \int \dfrac{ d l }{2 \pi} l C^{EE(\eta \eta^{'})}_l  \nonumber\\
& & \exp (-\frac{1}{2} l^2 [ \dfrac{C^{\eta\eta}_{gl}(0) + C^{\eta^{'}\eta^{'}}_{gl}(0)}{2} - C^{\eta\eta^{'}}_{gl}(r) ])  \nonumber\\
& & [J_4(lr) + \frac{l^2}{4} C^{\eta\eta^{'}}_{gl,2}(r) (J_2(lr) + J_6(lr)) +\ldots ],
\end{eqnarray}
\begin{eqnarray}
\widetilde{\xi}_{X}(r) & \equiv &  \langle {\rm e}^{-2i\phi_{\mathbf r}} \widetilde{T}(\mathbf{\hat{n}}) \widetilde{P}(\mathbf{\hat{n}}^{'}) \rangle                           \nonumber\\
& = & \int^{\eta_0}_0\, d\eta \int^{\eta_0}_0\, d\eta^{'} \langle {\rm e}^{-2i\phi_{\mathbf r}} T (\eta,\mathbf{\hat{n}} + \mathbf{\alpha}(\eta,\mathbf n)) (P^* (\eta^{'},\mathbf{\hat{n}}^{'} + \mathbf{\alpha}(\eta^{'},\mathbf n^{'})))^* \rangle          \nonumber\\
& = & \int^{\eta_0}_0\, d\eta \int^{\eta_0}_0\, d\eta^{'} \int \dfrac{d^2 \mathbf{l} }{(2 \pi)^2} \int \dfrac{d^2 \mathbf{l^{'}} }{(2 \pi)^2} \langle {\rm e}^{i \mathbf l \cdotp (\mathbf{\hat{n}} +\mathbf{\alpha} (\eta, \mathbf n))}                          \nonumber\\
& & {\rm e}^{- i \mathbf{l^{'}} \cdotp (\mathbf{\hat{n}}^{'} +\mathbf{\alpha} (\eta^{'}, \mathbf n^{'}))} \rangle {\rm e}^{-2i\phi_{\mathbf r}} {\rm e}^{ 2 i \phi_{\mathbf l^{'}} } (-1) \langle T(\eta,\mathbf l) E^{*}(\eta^{'},\mathbf l^{'}) \rangle   \nonumber\\
& = & \int^{\eta_0}_0\, d\eta \int^{\eta_0}_0\, d\eta^{'} \int \dfrac{d^2 \mathbf{l} }{(2 \pi)^2}  {\rm e}^{i \mathbf{l} \cdotp \mathbf{r} } C^{TE(\eta \eta^{'})}_l  {\rm e}^{2i\phi}(-1)  \nonumber\\
& &\exp (-\frac{1}{2} l^2 [ \dfrac{C^{\eta\eta}_{gl}(0) + C^{\eta^{'}\eta^{'}}_{gl}(0)}{2} - C^{\eta\eta^{'}}_{gl}(r) + \cos 2\phi C^{\eta\eta^{'}}_{gl,2}(r)  ])     \nonumber\\
& = & \int^{\eta_0}_0\, d\eta \int^{\eta_0}_0\, d\eta^{'} \int \dfrac{ d l }{2 \pi} l C^{TE(\eta \eta^{'})}_l  \nonumber\\
& & \exp (-\frac{1}{2} l^2 [ \dfrac{C^{\eta\eta}_{gl}(0) + C^{\eta^{'}\eta^{'}}_{gl}(0)}{2} - C^{\eta\eta^{'}}_{gl}(r) ])  \nonumber\\
& & [J_2(lr) + \frac{l^2}{4} C^{\eta\eta^{'}}_{gl,2}(r) (J_0(lr) +
J_4(lr)) +\ldots ].
\end{eqnarray}
In deriving the above three formulas we have ignored the $\langle T
\Psi \rangle$ and $\langle E \Psi \rangle$ terms, and used the
relationship $\phi = \phi_{\mathbf l} - \phi_{\mathbf r}$, $\langle
E(\eta,\mathbf{l}) E^*(\eta^{'},\mathbf{l^{'}}) \rangle = (2 \pi) ^2
\delta^2(\mathbf l - \mathbf l^{'}) C^{EE(\eta \eta^{'})}_l$ and
$\langle T(\eta,\mathbf{l}) E^*(\eta^{'},\mathbf{l^{'}}) \rangle =
(2 \pi) ^2 \delta^2(\mathbf l - \mathbf l^{'}) C^{TE(\eta
\eta^{'})}_l$. As in the temperature case, when all the deflection
angles $\mathbf{\alpha}(\eta, \mathbf{n})$ are set to
$\mathbf{\alpha}(\eta^{*}, \mathbf{n})$, the three polarization
correlators also reduce to the ordinary ones.

In the above we derived the four correlators
$\widetilde{\xi}(r),\widetilde{\xi}_{+}(r),\widetilde{\xi}_{-}(r),\widetilde{\xi}_{X}(r)$,
from which one obtains the power spectra
\begin{eqnarray}
C_l^T & = & 2 \pi \int r dr\,J_0(lr) \widetilde{\xi}(r)  \nonumber\\
C_l^E + C_l^B & = & 2 \pi \int r dr\,J_0(lr) \widetilde{\xi}_+(r)  \nonumber\\
C_l^E - C_l^B & = & 2 \pi \int r dr\,J_4(lr) \widetilde{\xi}_-(r)  \nonumber\\
C_l^{TE} & = & 2 \pi \int r dr\,J_2(lr) \widetilde{\xi}_X(r).
\end{eqnarray}

\section{Numerical results}
\label{sec:nr}
In this paper we use the Boltzmann code CMBFAST
\citep{Seljak:1996is} and adopt a flat $\Lambda CDM$ model with
cosmological parameters: $\Omega_b = 0.045$, $\Omega_c = 0.225$,
$\Omega_{\Lambda} = 0.73$, $H_0 = 70$, $\tau = 0.08$, $n_s = 1.0$
and $\Delta^2_{\zeta}=2.35 \times 10^{-10}$. When we do the integral
over $\eta$ and $\eta^{'}$, $C^{\psi\psi(\eta \eta^{'})}_l$ is
calculated as follows: 27 points $\eta_{(i)}$ are chosen in the
region $[\eta_{st}, \eta_0]$ ($\eta_{(1)} = \eta_{st}$, $\eta_{(27)}
= \eta_0$), where $\eta_{st}$ is the starting time of $\eta$ which
is chosen early enough to ensure the contribution before it can be
ignored due to the large optical depth, the same is done for
$\eta^{'}$; at these $27 \times 27$ places
$(\eta_{(i)},\eta^{'}_{(j)})$, $C^{\psi\psi(\eta \eta^{'})}_l$ is
directly calculated by Eq.(\ref{c12}); at the places where $\eta$
and $\eta^{'}$ are around the recombination era, even more
$C^{\psi\psi(\eta \eta^{'})}_l$ are directly calculated by
Eq.(\ref{c12}); at other places $C^{\psi\psi(\eta \eta^{'})}_l$ is
obtained by interpolation. Since $C^{\psi\psi(\eta \eta^{'})}_l$
vary slowly with time, the above approximation is good enough for
our purpose and simplify the numerical calculations.

\begin{figure}
\begin{center}
\includegraphics[width=0.6\textwidth]{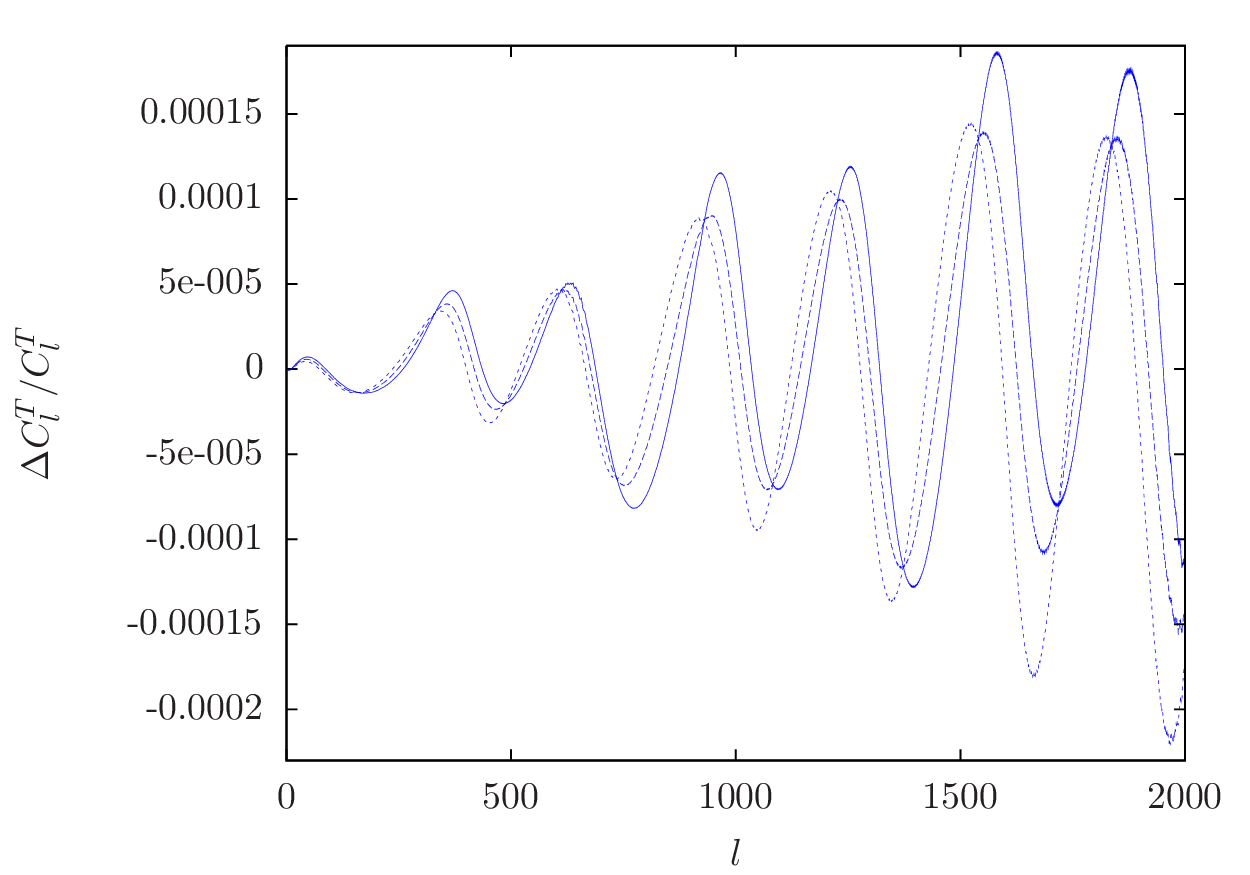}
\end{center}
\caption{Dependence of $\Delta C^T_l / C^T_l$ on $l$. $\Delta C^T_l$ is the power spectrum with the deflection angle fixed at $\eta^{fix}$ (computed by the method presented in \S 4.2 of \citep{Lewis:2006fu}) minus the one calculated in our paper, while $C^T_l$ is the unlensed power spectrum. The solid line, long dashed line and short dashed line correspond to $\eta^{fix}=270.6 Mpc,\eta^{fix}=285.4Mpc,\eta^{fix}=300.1Mpc$ respectively.}
\label{f2}
\end{figure}

\begin{figure}
\begin{center}
\includegraphics[width=0.6\textwidth]{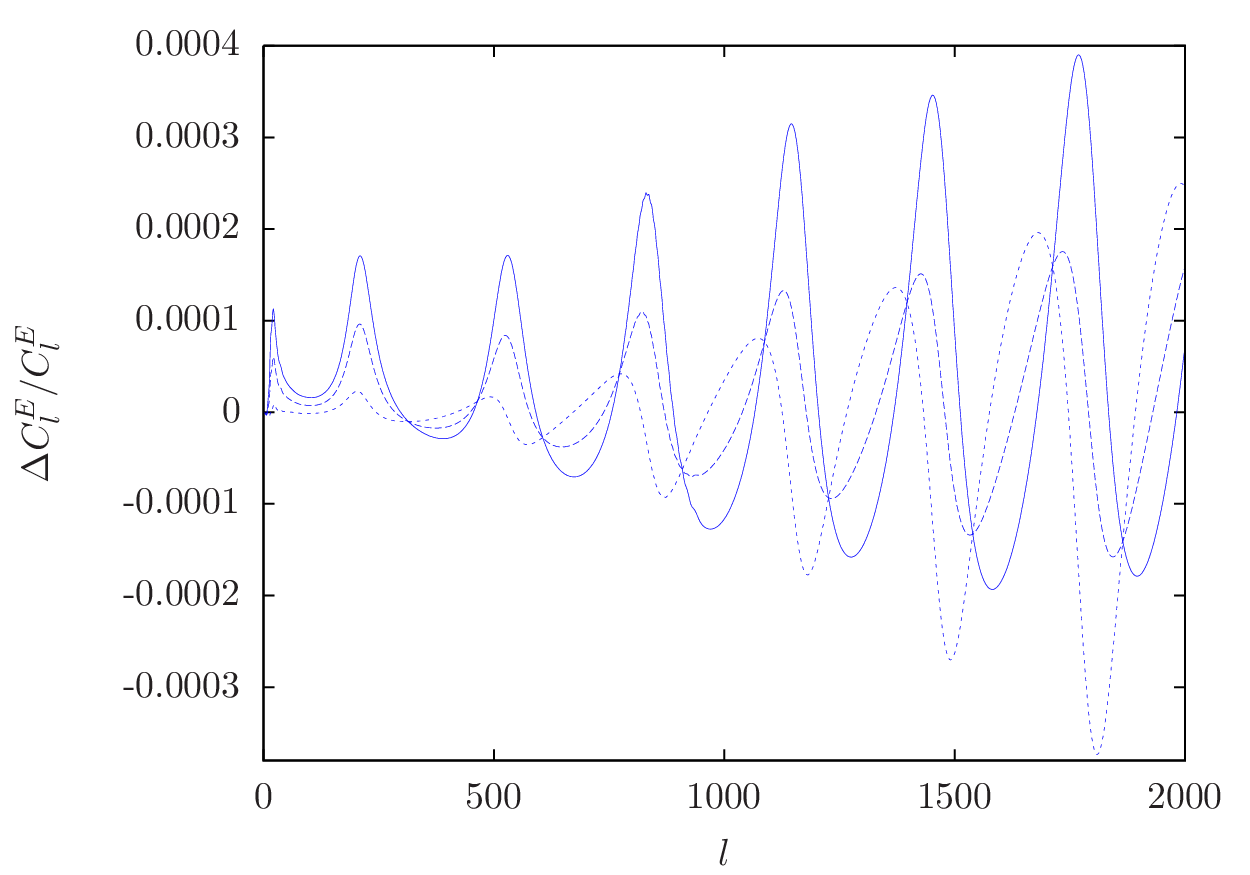}
\end{center}
\caption{Dependence of $\Delta C^E_l / C^E_l$ on $l$. $\Delta C^E_l$ is the power spectrum with the deflection angle fixed at $\eta^{fix}$ (computed by the method presented in \S 4.2 of \citep{Lewis:2006fu}) minus the one calculated in our paper, while $C^E_l$ is the unlensed power spectrum. The solid line, long dashed line and short dashed line correspond to $\eta^{fix}=270.6 Mpc,\eta^{fix}=285.4Mpc,\eta^{fix}=300.1Mpc$ respectively.}
\label{f3}
\end{figure}

\begin{figure}
\begin{center}
\includegraphics[width=0.6\textwidth]{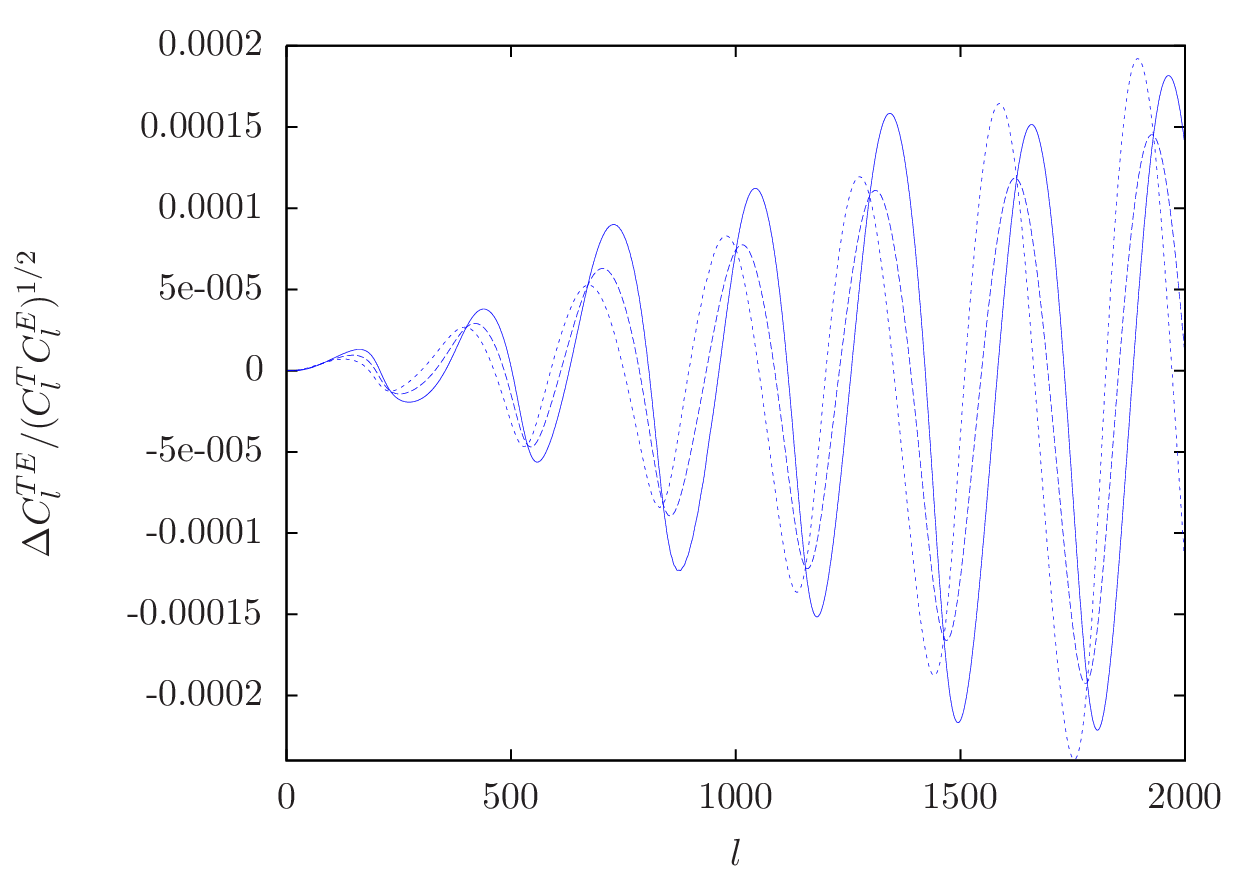}
\end{center}
\caption{Dependence of $\Delta C^{TE}_l / ({C^{T}_l C^{E}_l})^{1/2}$ on $l$. $\Delta C^{TE}_l$ is the power spectrum with the deflection angle fixed at $\eta^{fix}$ (computed by the method presented in \S 4.2 of \citep{Lewis:2006fu}) minus the one calculated in our paper, while $C^{T,E}_l$ is the unlensed power spectra. The solid line, long dashed line and short dashed line correspond to $\eta^{fix}=270.6 Mpc,\eta^{fix}=285.4Mpc,\eta^{fix}=300.1Mpc$ respectively.}
\label{f4}
\end{figure}

\begin{figure}
\begin{center}
\includegraphics[width=0.6\textwidth]{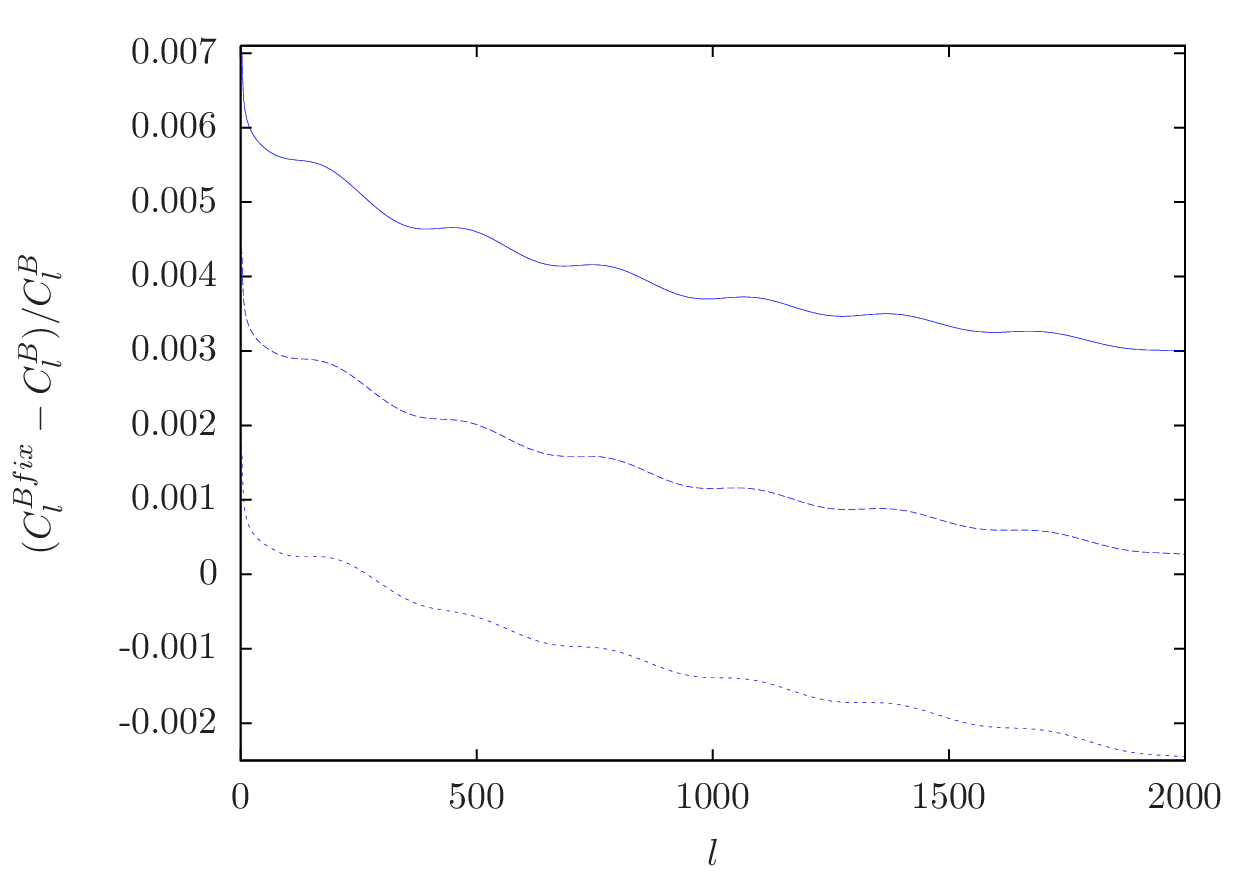}
\end{center}
\caption{Dependence of $(C^{Bfix}_l - C^{B}_l) / C^{B}_l$ on $l$.
$C^{Bfix}_l$ is the power spectrum with the deflection angle fixed
at $\eta^{fix}$ (computed by the method presented in \S 4.2 of
\citep{Lewis:2006fu}), while $C^{B}_l$ is the power spectrum
calculated in our paper in the cosmological model with reionization.
The solid line, long dashed line and short dashed line correspond to
$\eta^{fix}=270.6 Mpc,\eta^{fix}=285.4Mpc,\eta^{fix}=300.1Mpc$
respectively.} \label{f5}
\end{figure}

\begin{figure}
\begin{center}
\includegraphics[width=0.6\textwidth]{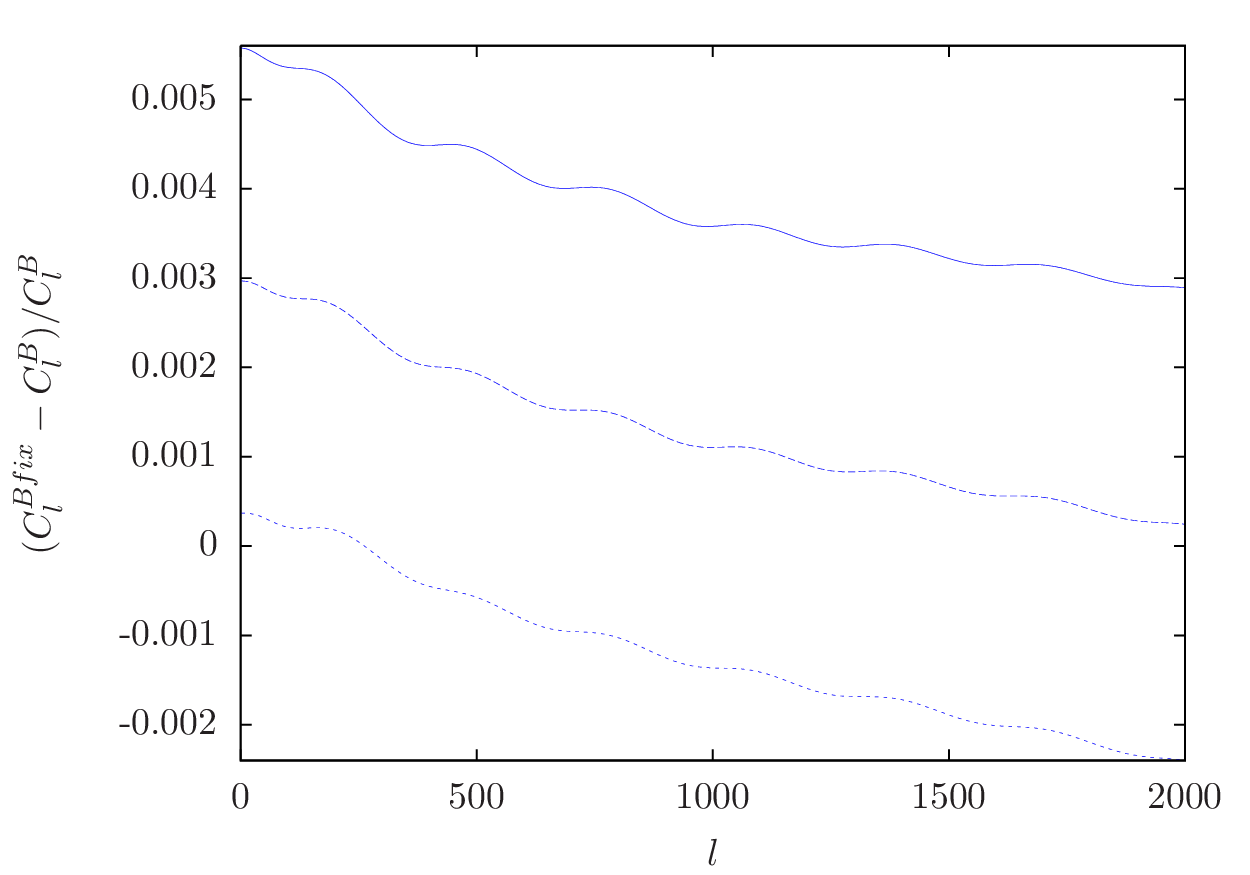}
\end{center}
\caption{Dependence of $(C^{Bfix}_l - C^{B}_l) / C^{B}_l$ on $l$.
$C^{Bfix}_l$ is the power spectrum with the deflection angle fixed
at $\eta^{fix}$ (computed by the method presented in \S 4.2 of
\citep{Lewis:2006fu}), while $C^{B}_l$ is the power spectrum
calculated in our paper in the cosmological model without
reionization. The solid line, long dashed line and short dashed line
correspond to $\eta^{fix}=270.6
Mpc,\eta^{fix}=285.4Mpc,\eta^{fix}=300.1Mpc$ respectively.}
\label{f6}
\end{figure}

In order to reflect the new effect, we also calculate the lensed
power spectra using the method presented in \S 4.2 and \S 5.3.2 of
\citep{Lewis:2006fu} (where the flat sky approximation  and
correlation function method are used but all the deflection angles
$\alpha(\eta, \mathbf{n})$ are set to $\alpha(\eta^{fix},
\mathbf{n})$). Here $\eta^{fix}$ is fixed at three values
$\eta^{fix}=270.6 Mpc,285.4Mpc,300.1Mpc$. The three times are around
the decoupling era and at $\eta^{fix}=285.4Mpc$ the visibility
function reaches its maximum. The final numerical results are
plotted in Figs \ref{f2}, \ref{f3}, \ref{f4}, \ref{f5}, \ref{f6}. In
Figs \ref{f2}, \ref{f3}, \ref{f4}, $\Delta C_{l}^{T,E,TE}$ is
defined as the power spectrum calculated at the deflection angle
fixed at $\eta^{fix}$ minus the power spectrum calculated in our
paper and $C_{l}^{T,E,TE}$ is the unlensed power spectrum. In Figs
\ref{f5}, \ref{f6}, $C_{l}^{Bfix}$ is the B mode power spectrum with
the deflection angle fixed at $\eta^{fix}$ and $C_{l}^{B}$ is the
one calculated in our paper.

The numerical results have some features. The first is that whatever
$\eta^{fix}$ is, one cannot completely reproduce the real case, and
generally speaking the spectra with $\eta^{fix}=285.4Mpc$ are the
best approximation which also deviate from the real case. For
$\eta^{fix}=285.4Mpc$, the B mode power spectrum change is of the
order $0.1 \%$, while other power spectra corrections are of the
order of $0.01 \%$. Simply speaking the effect on other types of
power spectra is about $10 \%$ of that on the B mode power spectrum.
This is what is we expected because the other power spectra
corrections due to weak gravitational lensing are of the order of
$10 \%$ at around $l=2000$ compared with the total power spectra.
The second is that the B mode power spectrum correction always
decrease. This is because the larger is $l$, the more contribution
comes from the early time.  The highest relative change of B mode
power spectrum minus the lowest one is about $0.4 \%$.  The third is
that the power spectra corrections (except the B mode) all intersect
at some $l$ approximately. We have checked this by fixing
$\eta^{fix}$ at some other times around the decoupling era and find
that this property still holds. Besides, we also calculated the B
mode power spectrum in the cosmological model with the same
parameters but without reionization (in this model $\eta^{fix}$ is
still fixed to be $\eta^{fix}=270.6 Mpc,285.4Mpc,300.1Mpc$ and at
$\eta^{fix}=285.4Mpc$ the visibility function still reaches its
maximum, see Fig 6). Comparing with the B mode power spectrum in the
cosmological model with reionization, one can find that the B mode
power spectrum in the cosmological model without reionization do not
drop fast at large scale, which indicates that the fast drop of B
mode power spectrum at the large scale in the reionization case is
indeed produced by reionization.


\section{Discussions}
\label{sec:dis} In this paper we consider that the deflection angle
of CMB photons varies with time.  Such an effect will affect the CMB
angular power spectra, especially the polarization B mode power
spectrum, which changes about $0.4 \%$. The smaller is the tensor
scalar ratio, the more important is this effect. Even when the
tensor scalar ratio gets as large as $0.1$, the changes in B mode
power spectrum induced by this effect cannot be ignored at small
scales. If one want to use CMB data to provide stringent constraint
on the inflationary physics, for example the power spectrum index of
the primordial tensor fluctuation, this effect may need to be
considered.

\begin{acknowledgments}
We would like to thank Dr Sun Wei-Min and Qi Shi for improving the
manuscript.
\end{acknowledgments}

\appendix
\section{Correlator}
In this Appendix we calculate $C^{TT(\eta \eta^{'})}_l$, $C^{EE(\eta
\eta^{'})}_l$, $C^{TE(\eta \eta^{'})}_l$ in the case of scalar
primordial perturbations. First we note that
\begin{eqnarray}
T(\eta,\mathbf{n}) & = & \int \dfrac{d^3 \mathbf{k} }{(2 \pi)^3} \xi (\mathbf k) {\rm e}^{i \mathbf{k} \cdotp (\eta_0 - \eta) \mathbf{\hat{n}} } S_T(\eta,\mathbf{k}) \nonumber\\
& = & \int \dfrac{d^3 \mathbf{k} }{(2 \pi)^3} \xi (\mathbf k)  S_T(\eta,\mathbf k) 4 \pi \sum_{lm} i^l j_l(x) Y^*_{lm}(\mathbf{\hat{k}}) Y_{lm}(\mathbf{\hat{n}}).
\end{eqnarray}
In deriving the last line we have made use of the formula ${\rm e}^{i \mathbf k \cdotp \mathbf x} = 4 \pi \sum_{lm} i^l j_l(kx)Y^*_{lm}(\mathbf{\hat{k}}) Y_{lm}(\mathbf{\hat{x}})$. With the above formula, it is easy to obtain the harmonic components of $T(\eta,\mathbf{n})$
\begin{eqnarray}
T_{lm}(\eta) = \int \dfrac{d^3 \mathbf{k} }{(2 \pi)^3} \xi (\mathbf k) S_T(\eta,\mathbf{k}) 4 \pi i^l j_l(x) Y^{*}_{lm}(\mathbf{\hat{k}}).
\end{eqnarray}
After performing the angular integral, one obtains
\begin{eqnarray}
\langle T_{lm}(\eta) T_{l^{'}m^{'}}(\eta^{'})\rangle = \delta_{ll^{'}} \delta_{mm^{'}} C^{TT(\eta \eta^{'})}_l
\end{eqnarray}
with
\begin{eqnarray}
C^{TT(\eta \eta^{'})}_l = \int k^2 dk \frac{2}{\pi} P_{\xi}(k)  S_T(\eta,k) j_l(x) S_T(\eta^{'},k) j_l(x^{'}),
\end{eqnarray}
where $\langle \xi (\mathbf{k}) \xi^*(\mathbf{k}^{'}) \rangle = (2 \pi)^3 P_{\xi}(k) \delta^3(\mathbf{k} - \mathbf{k}^{'})$ has been used. In the flat sky approximation it becomes
\begin{eqnarray}
\langle T(\eta,\mathbf{l}) T^*(\eta^{'},\mathbf{l^{'}}) \rangle = (2 \pi) ^2 \delta^2(\mathbf l - \mathbf l^{'}) C^{TT(\eta \eta^{'})}_l.
\end{eqnarray}
From Eq.(\ref{psi}) one have
\begin{equation}
\psi(\eta,\mathbf n) = -2 \int_0^{\chi} d\chi_1 \frac{f_K(\chi-\chi_1)}{f_K(\chi)f_K(\chi_1)} \int \dfrac{d^3 \mathbf{k} }{(2 \pi)^3} \xi (\mathbf k) {\rm e}^{i \mathbf{k} \cdotp (\eta_0 - \eta_1) \mathbf{\hat{n}} } T_{\Psi}(\eta_1,\mathbf{k}),
\end{equation}
where $T_{\Psi}(\eta_1,\mathbf{k})$ is the transfer function of 'Weyl potential'. Since $\Psi = \frac{(\Psi_N - \Phi_N)}{2}$, $T_{\Psi}(\eta_1,\mathbf{k})$ can be directly computed by the CMBFAST code \citep{Seljak:1996is}. Following the same steps as above, one obtains
\begin{eqnarray}
\langle \psi(\eta,\mathbf{l}) \psi^*(\eta^{'},\mathbf{l^{'}}) \rangle = (2 \pi) ^2 \delta^2(\mathbf l - \mathbf l^{'}) C^{\psi\psi(\eta \eta^{'})}_l
\end{eqnarray}
with
\begin{eqnarray}
C^{\psi\psi(\eta \eta^{'})}_l & = & 4 \int_0^{\chi} d\chi_1 \dfrac{\chi-\chi_1}{\chi \chi_1} \int_0^{\chi^{'}} d\chi_1^{'} \dfrac{\chi^{'}-\chi_1^{'}}{\chi^{'} \chi_1^{'}}     \nonumber \\
& &\int k^2 dk \frac{2}{\pi} P_{\xi}(k)  T_{\Psi}(\eta_1,k) j_l(x_1)
T_{\Psi}(\eta_1^{'},k) j_l(x_1^{'}). \label{c12}
\end{eqnarray}
From Eq.(\ref{ebar}) one have
\begin{eqnarray}
\bar{E}(\eta,\mathbf{n}) & = & \int\,\dfrac{d^3 \mathbf{k} }{(2 \pi)^3} \xi({\mathbf k})\, \frac{3}{4} g(\eta)\,\Pi(\eta,k)
        (1+\partial_x^2)^2 \left( x^2\,{\rm e}^{-ix\mu}\right).
\end{eqnarray}
After doing the same thing, one can get the following formulas
\begin{eqnarray}
\langle \bar{E}(\eta,\mathbf{l}) \bar{E}^*(\eta^{'},\mathbf{l^{'}}) \rangle & = & (2 \pi) ^2 \delta^2(\mathbf l - \mathbf l^{'}) C^{\bar{E}\bar{E}(\eta \eta^{'})}_l  \nonumber\\
\langle E(\eta,\mathbf{l}) E^*(\eta^{'},\mathbf{l^{'}}) \rangle & = & (2 \pi) ^2 \delta^2(\mathbf l - \mathbf l^{'}) S^2_l C^{\bar{E}\bar{E}(\eta \eta^{'})}_l  \nonumber\\
& = &  (2 \pi) ^2 \delta^2(\mathbf l - \mathbf l^{'}) C^{EE(\eta \eta^{'})}_l
\end{eqnarray}
with $S_l = [(l-1)l(l+1)(l+2)]^{-1/2}$ and
\begin{eqnarray}
C^{EE(\eta \eta^{'})}_l = S^2_l \int k^2 dk \frac{2}{\pi} P_{\xi}(k)
\frac{3}{4} g(\eta) \Pi(\eta,k) \dfrac{j_l(x)}{x^2} \frac{3}{4}
g(\eta^{'}) \Pi(\eta^{'},k) \dfrac{j_l(x^{'})}{x^{'2}}.
\end{eqnarray}
For cross terms we also have
\begin{eqnarray}
\langle T(\eta,\mathbf{l}) \bar{E}^*(\eta^{'},\mathbf{l^{'}}) \rangle & = & (2 \pi) ^2 \delta^2(\mathbf l - \mathbf l^{'}) C^{T\bar{E}(\eta \eta^{'})}_l  \nonumber\\
\langle T(\eta,\mathbf{l}) E^*(\eta^{'},\mathbf{l^{'}}) \rangle & = & (2 \pi) ^2 \delta^2(\mathbf l - \mathbf l^{'}) S_l C^{T\bar{E}(\eta \eta^{'})}_l  \nonumber\\
& = &  (2 \pi) ^2 \delta^2(\mathbf l - \mathbf l^{'}) C^{TE(\eta \eta^{'})}_l
\end{eqnarray}
with
\begin{eqnarray}
C^{TE(\eta \eta^{'})}_l = S_l \int k^2 dk \frac{2}{\pi} P_{\xi}(k)
S_T(\eta,k) j_l(x) \frac{3}{4} g(\eta^{'}) \Pi(\eta^{'},k)
\dfrac{j_l(x^{'})}{x^{'2}}.
\end{eqnarray}

\bibliographystyle{apj}

%\bibliography{rf.bib}

\end{document}